\documentclass[5p]{elsarticle}

\usepackage{epsfig}
\usepackage{graphicx}
\usepackage{bm}

\begin{document}
\title{The problem of repulsive quark interactions - Lattice versus mean field models}


\author[itp,fias]{J.~Steinheimer}
\ead{steinheimer@th.physik.uni-frankfurt.de}
\author[itp,fias,csc]{S.~Schramm}
\address[itp]{Institut f\"ur Theoretische Physik, Goethe-Universit\"at, Max-von-Laue-Str.~1,
D-60438 Frankfurt am Main, Germany}
\address[fias]{Frankfurt Institute for Advanced Studies (FIAS), Ruth-Moufang-Str.~1, D-60438 Frankfurt am Main,
Germany}
\address[csc]{Center for Scientific Computing, Max-von-Laue-Str.~1, D-60438 Frankfurt am Main}

\begin{abstract}
We calculate the 2nd and 4th order quark number susceptibilities at zero baryochemical potential, using a PNJL approach and an approach which includes, in a single model, quark and hadronic degrees of freedom.
We observe that the susceptibilities are very sensitive to possible quark-quark vector interactions. Compared to lattice data our results suggest that above $T_c$ any mean field type of repulsive vector interaction can be excluded from model calculations.
Below $T_c$ our results show only very weak sensitivity on the strength of the quark and hadronic vector interaction. The best description of lattice data around $T_c$ is obtained for a case of coexistence of hadronic and quark degrees of freedom .
\end{abstract}

\maketitle

Many recent experimental programs in heavy ion physics, e.g. at the relativistic heavy ion collider and the planned FAIR facility, are aimed at a better understanding of bulk properties of QCD matter. In particular the 
experimental confirmation of the deconfinement and chiral phase transitions as well as the search for the critical end point (CeP) are of great interest to the community.
Lattice results at $\mu_B = 0$ suggest that both phase transitions exhibit a smooth crossover. Suffering from the so called sign problem at finite baryo chemical potential, lattice studies have so far not been able to 
constitute a consistent picture of the phase diagram and the existence of the CeP \cite{Fodor:2004nz,deForcrand:2007rq}. While a Taylor expansion of lattice results at $\mu_b=0$ predict the existence of a critical end point these results may strongly depend on the order to which the Taylor coefficients can be evaluated \cite{Klein:2010tk}. Lattice calculations at imaginary chemical potential on the other hand show no indication for a CeP, though these calculations are done on on a coarse lattice with a fermion action that is known to have large discretization errors. Other studies trying to locate the CeP often rely on the applicability of various effective models where CeP's existence and location seems very sensitive on the vector coupling strength at and below $T_c$ \cite{Fukushima:2008is}.

\section{The PNJL model}
The PNJL model was introduced in \cite{Fukushima:2003fw,Ratti:2005jh} as an effective chiral quasi-quark model that incorporates a mean field like coupling to a color background field. It has often been shown to reproduce many general features of lattice results at $\mu_B=0$ \cite{Ratti:2004ra,Roessner:2006xn,Sasaki:2006ww,Ratti:2007jf,Rossner:2007ik,Ciminale:2007sr,Schaefer:2007pw,Fu:2007xc,Hell:2008cc,Abuki:2008nm,Fukushima:2008wg,Fukushima:2008is,Costa:2008gr,Costa:2008dp,Hansen:2006ee,Mukherjee:2006hq,Abuki:2008tx,Abuki:2008ht,Fukushima:2009dx,Mao:2009aq,Schaefer:2009ui,Ghosh:2006qh,Karsch:2010hm}.
In our comparison we will use a very basic parametrization of the two-flavor PNJL model and extend it to incorporate a repulsive vector interaction.
The thermodynamic potential of our parametrization reads:
\begin{equation}
\Omega = U(\Phi,\Phi^*,T)+\sigma^2/2 G_S-\omega^2/2 G_V-\Omega_{q}
\end{equation} 
with
\begin{eqnarray}
\Omega_{q}&=&2 N_f \int \frac{d^3 p}{(2 \pi)^3}\nonumber \\
&&\left\{ T \ln\left[1+3\Phi e^{-(E_p-\mu_q^*)/T}\right.\right. \nonumber \\
&+&\left.3\Phi^* e^{-2(E_p-\mu_q^*)/T}+e^{-3(E_p-\mu_q^*)/T}\right] \nonumber \\
&+&T \ln\left[1+3\Phi^* e^{-(E_p+\mu_q^*)/T} \right.\nonumber \\
&+&3\Phi e^{-2(E_p+\mu_q^*)/T}\left.e^{-3(E_p+\mu_q^*)/T}\right] \nonumber \\
&+&\left. 3 \Delta E_p \Theta(\Lambda^2-\vec{p} ^2)\right\}
\end{eqnarray}
Here the grand canonical potential includes contributions of sates with 1, 2 and 3 times the single quark energy. Note that the 3 quark state does not couple to the Polyakov loop.\\
The (traced) Polyakov loop $\Phi$ was introduced as:
\begin{equation}
\Phi= 1/3 \ Tr \ e^{i \phi/T}
\end{equation}
with $\phi=A_4$, a background color gauge field. The thermodynamics of $\Phi$ (and $\Phi^*$) are controlled by the effective potential $U(\Phi,\Phi^*,T)$ \cite{Ratti:2006wg}:\\
\begin{eqnarray}\label{polyakov} 
	U&=&-\frac12 a(T)\Phi\Phi^* \nonumber \\ 
	&+&b(T)\ln[1-6\Phi\Phi^*+4(\Phi^3\Phi^{*3})-3(\Phi\Phi^*)^2]
\end{eqnarray} 
 with 
 $a(T)=a_0 T^4+a_1 T_0 T^3+a_2 T_0^2 T^2$, \ $b(T)=b_3 T_0^3 T$.\\

This choice of effective potential satisfies the $Z(3)$ center symmetry of the pure gauge Lagrangian. In the confined phase, $U$ has a minimum at $\Phi=0$, while above the critical Temperature $T_0$ its minimum is shifted to finite values of $\Phi$. The logarithmic term appears from the Haar measure of the group integration with respect to the SU(3) Polyakov loop matrix. The parameters $a_0, a_1, a_2$ and $b_3$ are fixed, as in \cite{Ratti:2006wg}, by demanding a first order phase transition in the pure gauge sector at $T_0=270MeV$, and that the Stefan-Boltzmann limit is reached for $T \rightarrow \infty$.

The dynamical mass of the quarks $m=m_0- \sigma= m_0 - G_S \left\langle \overline{\Psi} \Psi \right\rangle$ is the same as in the NJL model and the vector coupling induces an effective chemical potential for the quarks $\mu_q^*=\mu_q + \omega=\mu_q + G_V \left\langle \Psi^{\dagger} \Psi \right\rangle$. The two auxiliary fields $\sigma$ and $\omega$ are controlled by the potential terms and the last term includes the difference $\Delta E_p$ between the quasi particle energy and the energy of free quarks. The NJL part of the model has 4 parameters, the bare quark mass for the $u$- and $d$-quarks (assuming isospin symmetry), the three-momentum cutoff of the quark-loop integration  $\lambda$ and the coupling strengths $G_S$ and $G_V$. To reproduce realistic values for the pion mass and decay constant as well as the chiral condensate, we take these values to be \cite{Ratti:2005jh}: 
 $m_{u,d}=5.5 \ \rm{MeV} ,G_S=10.08 \ \rm{GeV}^{-2}, \Lambda=651 \ \rm{MeV}$ ($G_V$ will be left as a model parameter to study the influence of the vector coupling on our results).\\
The self consistent solutions are obtained by minimizing the thermodynamic potential with respect to the fields $\sigma$, $\omega$, $\Phi$ and $\Phi^*$.

\section{The QH model}
To estimate the influence, of hadronic contributions, on the susceptibilities we will compare the results from the PNJL model with those obtained from a model where quark and hadronic degrees of freedom are present in a single partition function (Quark-Hadron model).\\
In the following we will shortly describe the different components of the model, for a more detailed discussion we refer to \cite{Steinheimer:2010ib}.
The hadronic part is described by a flavor-SU(3) model, which
is an extension of a non-linear representation of a $\sigma$-$\omega$ model including
the pseudo-scalar and vector nonets of mesons and the baryonic octet \cite{Papazoglou:1998vr,Papazoglou:1997uw,Dexheimer:2008ax}.
Besides the kinetic energy term for hadrons and quarks, the terms:
\begin{eqnarray}
&L_{int}=-\sum_i \bar{\psi_i}[\gamma_0(g_{i\omega}\omega+g_{i\phi}\phi)+m_i^*]\psi_i,&\\
&L_{meson}=-\frac{1}{2}(m_\omega^2 \omega^2+m_\phi^2\phi^2)&\nonumber\\
&-g_4\left(\omega^4+\frac{\phi^4}{4}+3\omega^2\phi^2+\frac{4\omega^3\phi}{\sqrt{2}}+\frac{2\omega\phi^3}{\sqrt{2}}\right)\nonumber&\\
&+\frac{1}{2}k_0(\sigma^2+\zeta^2)-k_1(\sigma^2+\zeta^2)^2-k_2\left(\frac{\sigma^4}{2}+\zeta^4\right)&\nonumber\\
&-k_3\sigma^2\zeta+ m_\pi^2 f_\pi\sigma+\left(\sqrt{2}m_k^ 2f_k-\frac{1}{\sqrt{2}}m_\pi^ 2 f_\pi\right)\zeta &\nonumber\\
&+ \chi^4-\chi_0^4 + \ln\frac{\chi^4}{\chi_0^4} -k_4\ \frac{\chi^4}{\chi_0^4} \ln{\frac{\sigma^2\zeta}{\sigma_0^2\zeta_0}} ~.&
\label{formel2}
\end{eqnarray}
represent the interactions between baryons
and vector and scalar mesons, the self-interactions of
scalar and vector mesons, and an explicitly chiral symmetry breaking term.
The index $i$ denotes the baryon octet and the different quark flavors. Here, the mesonic condensates (determined in
mean-field approximation) included are
the vector-isoscalars $\omega$ and $\phi$, and
the scalar-isoscalars $\sigma$ and $\zeta$ (strange quark-antiquark state).
The last four terms of eqn.(\ref{formel2}) were introduced to model the QCD trace anomaly, where the dilaton field $\chi$ can be identified with the gluon condensate.

The effective masses of the baryons and quarks
are generated by the scalar mesons except for an explicit
mass term ($\delta m_N=120$ MeV, $\delta m_q=5$ MeV and $\delta m_s=105$ MeV for the strange quark), 
$ m_{i}^*=g_{i\sigma}\sigma+g_{i\zeta}\zeta+\delta m_i.$\\
Vector type interactions introduce an effective chemical potential for the quarks and baryons, generated by the coupling to the vector mesons: $\mu^*_i=\mu_i-g_{i\omega}\omega-g_{i\phi}\phi$.\\
The coupling constants for the baryons \cite{Dexheimer:2009hi} are chosen
to reproduce the vacuum masses of the baryons, nuclear saturation properties and
asymmetry energy as well as the $\Lambda$-hyperon optical potential. The vacuum expectation values of the scalar
mesons are constrained by reproducing the pion and kaon decay constants. For the quarks we chose the following coupling parameters: $g_{q\sigma}=g_{s\zeta}= 4.0$, while the vector coupling strength $g_{q\omega}$ is left as free parameter.
The coupling of the quarks to the Polyakov loop is introduced in the thermal energy of the quarks \cite{Steinheimer:2010ib}.
All thermodynamical quantities, energy density $e$, entropy density $s$ as well as the
densities of the different particle species $\rho_i$, can be derived from the grand canonical potential:
\begin{equation}
\Omega=-L_{int}-L_{meson}+\Omega_{th}-U
\end{equation}
Here $\Omega_{th}$ includes the heat bath of hadronic and quark quasi particles. The effective potential $U(\Phi,\Phi^*,T)$ which controls the dynamics of the Polyakov-loop has the form of eqn.(\ref{polyakov}).\\
Since we expect the hadronic contribution to disappear, at least at some point above $T_c$, we included
effects of finite-volume particles to effectively suppresses the hadronic degrees of freedom, when deconfinement is achieved.
Including these effects in a thermodynamic model for hadronic matter, was proposed some time ago \cite{Hagedorn:1980kb,Baacke:1976jv,Gorenstein:1981fa,Hagedorn:1982qh}. We will use an ansatz similar to that in \cite{Rischke:1991ke,Cleymans:1992jz}, but modify it to also treat the point like quark degrees of freedom consistently.
We introduce the chemical potential $\widetilde{\mu}_i$ which is connected to the real chemical potential $\mu_i^*$, of the $i$-th particle species by the relation: $	\widetilde{\mu}_i=\mu_i^*-v_{i} \ P$
($P$ is the sum over all partial pressures and $v_i$ the volume of a hadron). All thermodynamic quantities can then be calculated with respect to the temperature $T$ and the new chemical potentials $\widetilde{\mu}_i$. To be thermodynamically consistent, all densities have to be multiplied by a volume correction factor $f$, which is the ratio of the total volume $V$ and the reduced volume $V'$, not being occupied.
In this configuration the chemical potentials of the hadrons are decreased by the quarks, but not vice versa. As the quarks start appearing they effectively suppress the hadrons by changing their chemical potential, while the quarks are only affected through the volume correction factor $f$. Our description of the excluded volume effects is admittedly simplified and parameter dependent, but it enables us to describe a phase transition from hadronic to quark degrees of freedom in a quite natural and thermodynamically consistent manner.\\

\begin{figure}[t]
\centering
\includegraphics[width=0.5\textwidth]{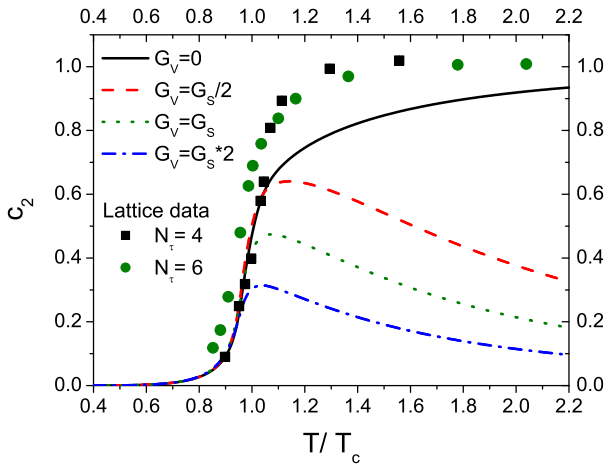}
\caption{\label{pnjlc2} The second order quark number susceptibility from the PNJL model, with different strengths of the vector interaction, as a function of $T$ over $T_c$.
Black solid line: no vector interaction, red dashed line: $G_V=G_S/2$, green dotted line: $G_V=G_S$, blue dash dotted line $G_V=2 G_S$.
}
\end{figure}

As has been mentioned above, the Lagrangian of the chiral model contains dilaton terms to model the scale anomaly. These terms constrain the chiral condensate, if the dilaton is frozen at its ground state value $\chi_0$. On the other hand, as deconfinement is realized, the expectation value of the chiral condensate should vanish at some point. On account of this we will couple the Polyakov loop to the dilaton in the following way:
\begin{equation}
	\chi=\chi_0 \ (1-0.5(\Phi\Phi^*))
\end{equation}
Assuming a hard part for the dilaton field which essentially stays unchanged and a soft part, which vanishes when deconfinement is realized. 
Hence, allowing the chiral condensate to also approach zero.\\
For a more detailed discussion of the model and comparisons with lattice data we refer to \cite{Steinheimer:2010ib}. 
\section{Susceptibilities}

Lattice results at finite chemical potentials are often obtained as Taylor expansion of the thermodynamic quantities in the parameter $\mu/T$ around zero chemical potential \cite{Allton:2002zi}. In the Taylor expansion of the pressure $p=-\Omega$, the coefficients, which can be identified with the quark number susceptibilities, follow from:
\begin{eqnarray}
\frac{p(T,\mu_B)}{T^4}&=& \sum_{n=0}^{\infty}c_n(T)\left(\frac{\mu_B}{T}\right)^n \\
c_n(T)&=&\left. \frac{1}{n!} \frac{\partial^n(p(T,\mu_B)/T^4)}{\partial(\mu_B/T)^n}\right|_{\mu_B=0}
\end{eqnarray}

\begin{figure}[t]
\centering
\includegraphics[width=0.5\textwidth]{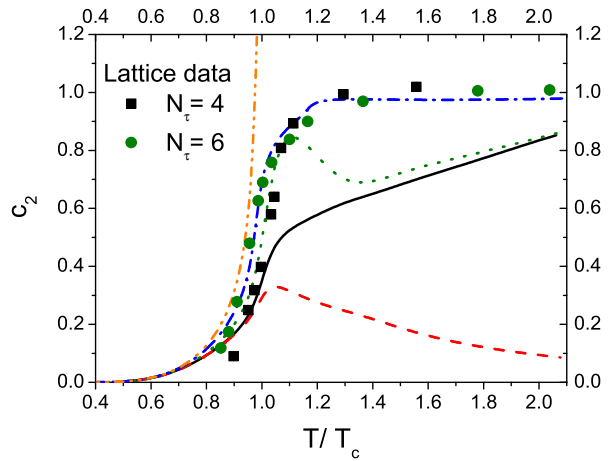}
\caption{\label{qhmc2}The second order quark number susceptibility from the QHM model, with different strengths of the vector interaction, as a function of $T$ over $T_c$.
Black solid line: $g_{N\omega}\ne 0, g_{q\omega}= 0, v=1$, red dashed line:  $g_{N\omega}\ne 0, g_{q\omega}= g_{N\omega}/3, v=1$,
green dotted line: $g_{N\omega}= 0, g_{q\omega}= 0, v=1$, blue dashed dotted line: $g_{N\omega}\ne 0, g_{q\omega}= 0, v=0$, orange dash dot dot line: $g_{N\omega}= 0, g_{q\omega}= 0, v=0$.
}
\end{figure} 

In our approach we explicitly calculate the pressure at finite $\mu_B$ and then extract the expansion coefficients numerically.
The results for the second coefficient calculated for the PNJL model and QH model, compared to lattice results \cite{Cheng:2008zh}, are shown in figures (\ref{pnjlc2}) and (\ref{qhmc2}). One can clearly observe that the best description can be obtained when quark vector interactions are turned off. Any form of repulsive interaction strongly decreases the value of the second order coefficient above $T_c$. The lattice results on the other hand quickly reach a value that is expected for a non-interacting gas of quarks, even right above $T_c$. This is in fact surprising as other thermodynamic quantities tend to favor a picture with a wide region around $T_c$ where interactions are strong. This behavior was also reproduced by the PNJL model and the QH model while both fail to describe the fast increase in $c_2$ right above $T_c$.\\

Next we try to disentangle the hadronic contribution to the second coefficient. As can be seen in figure (\ref{qhmc2}) the dependence on the strength of the repulsive interaction below $T_c$ is rather small and one can not exclude any scenario. In the crossover region, around $T_c$, differences become obvious.\\
The solid black line in Figure (\ref{qhmc2}) displays the result for $c_2$ using the standard parametrization of the QH model as described in \cite{Steinheimer:2010ib} without any repulsive quark-quark interactions. The value of $c_2$ only slowly approaches 1, which is mainly due to the fact that the value of the chiral condensate drops to 0 rather slow in our model and therefore the quark masses do not decrease as fast as in the PNJL approach.\\

Including repulsive vector interactions for the quarks gives a result which is similar to the one obtained from the PNJL (red dashed line). Here the repulsive interactions strongly decrease the value of $c_2$.\\

Turning off all repulsive interactions, vector interactions for quarks and hadrons as well as the excluded volume corrections (orange dash dot dotted line), leads to an drastic overestimation of $c_2$. This is expected as all hadronic degrees of freedom are present at and above $T_c$ if the excluded volume effects are turned off. Therefore one largely overestimates the effective degrees of freedom.\\

On the other hand if the repulsive vector interactions are turned on only for the hadrons one obtains a rather good description of the lattice results (blue dash dotted line). In this parametrization the hadrons are also still present in the system up to arbitrary high temperatures, as the excluded volume effects are still turned of, and therefore all thermodynamic quantities are largely over predicted.\\

To remove the hadronic contributions from the system we introduced excluded volume corrections as described above. This leads to a pronounced dip in $c_2$ above $T_c$ (green dotted line), indicating that our excluded volume approach is either to simplified or all hadronic contributions are already vanishing completely right above $T_c$ or that the lattice results are still not accurate enough to sufficiently resolve effects of hadron hadron interactions.\\

\begin{figure}[t]
\centering
\includegraphics[width=0.5\textwidth]{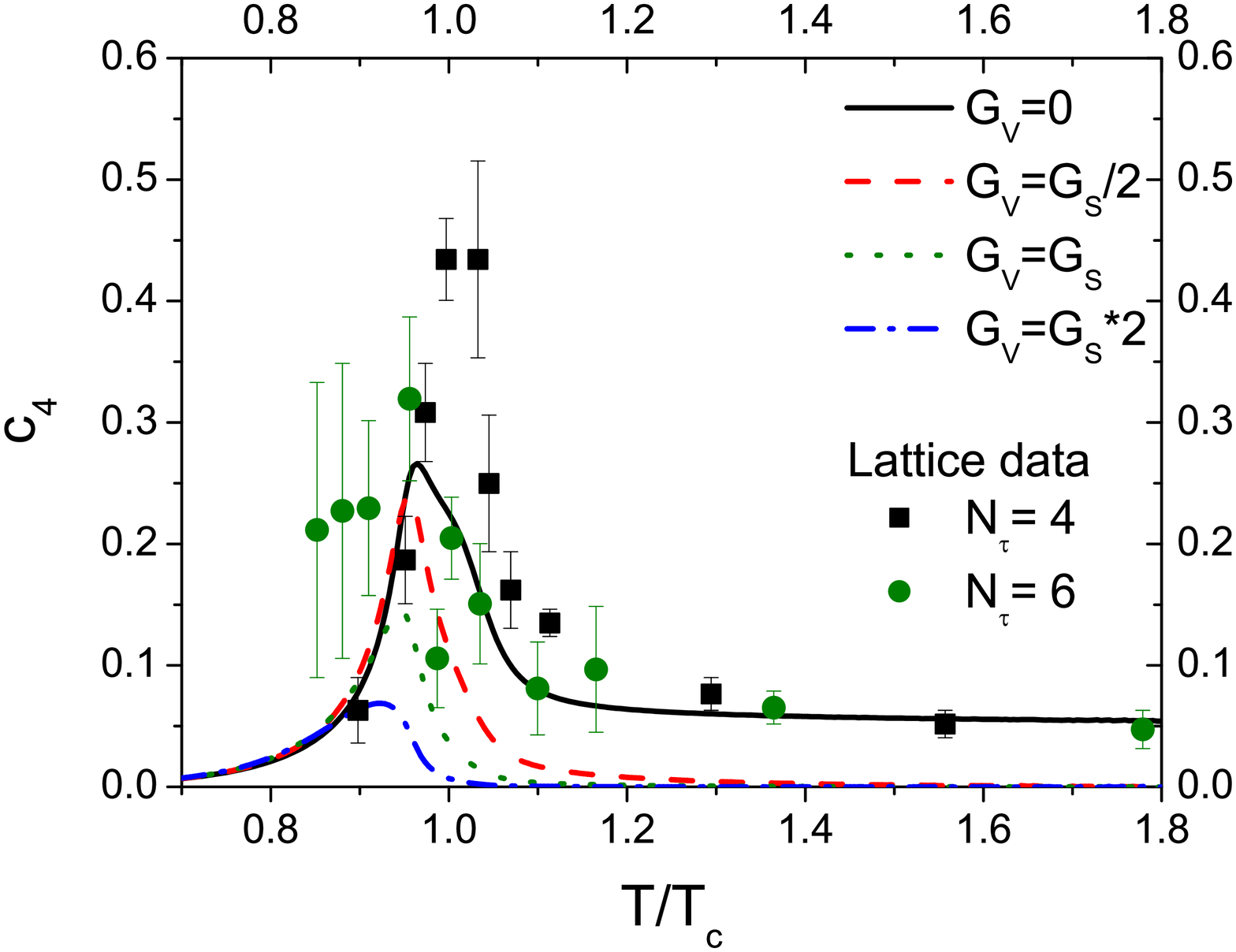}
\caption{\label{pnjlc4}The forth order quark number susceptibility from the PNJL model, with different strengths of the vector interaction, as a function of $T$ over $T_c$.
Black solid line: no vector interaction, red dashed line: $G_V=G_S/2$, green dotted line: $G_V=G_S$, blue dash dotted line: $G_V=2 G_S$. 
}
\end{figure}

\begin{figure}[t]
\centering
\includegraphics[width=0.5\textwidth]{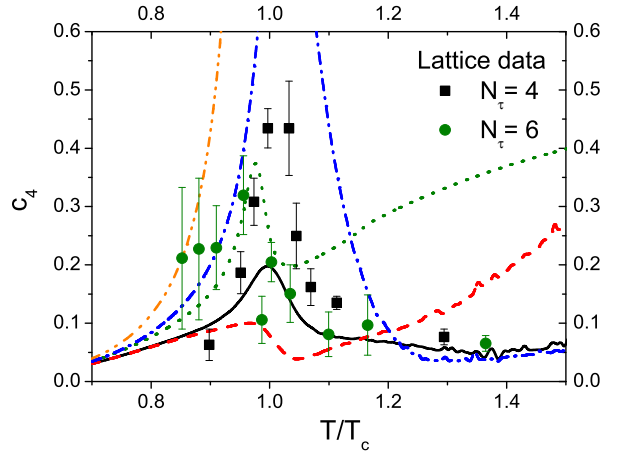}
\caption{\label{qhmc4} The fourth order quark number susceptibility from the QHM model, with different strengths of the vector interaction, as a function of $T$ over $T_c$.
Black solid line: $g_{N\omega}\ne 0, g_{q\omega}= 0, v=1$, red dashed line:  $g_{N\omega}\ne 0, g_{q\omega}= g_{N\omega}/3, v=1$,
green dotted line: $g_{N\omega}= 0, g_{q\omega}= 0, v=1$, blue dashed dotted line: $g_{N\omega}\ne 0, g_{q\omega}= 0, v=0$, orange dash dot dot line: $g_{N\omega}= 0, g_{q\omega}= 0, v=0$.
}
\end{figure} 

The fourth order coefficients calculated for the PNJL and QH model are shown in figures \ref{pnjlc4} and  \ref{qhmc4}. Although the errors on the lattice results are still significant our results for $c_4$ support the statements made for $c_2$. At temperatures above $T_c$ a gas of free non interacting quarks, without any hadronic contribution, gives the only reasonable description of the data already slightly above $T_c$. While the height of the peak does depend on the strength of the vector coupling, we observe also a strong dependence on the slope of the change of the order parameters around $T_c$, which is larger in the QH model as compared to the PNJL model. Notice that the parametrization shown as the blue dash-dotted line, which gave the best description for $c_2$, overestimates $c_4$ around $T_c$ drastically. This is simply because all thermodynamic quantities are over predicted in this case and therefore the densities as well as the order parameters increase steeper as they do in the lattice calculations. 

\section{Conclusion}
We have compared results for the quark number susceptibilities at $\mu_B=0$ obtained from two mean field models to recent results from lattice calculations. Both models strongly indicate that the EoS of QCD above $T_c$ seems to be composed purely of a gas  of non-interacting quasi particles. This finding is in agreement with recent work by \cite{Ferroni:2010xf} where the extracted quark vector coupling strength approaches zero very fast around $T_c$. On the other hand, lattice observables like the interaction measure and the normalized Polyakov loop indicate a large region above the critical temperature where the hot QCD medium is far from being an ideal gas.
Around $T_c$ repulsive hadronic interactions are supported by our results with the QH model. To describe the steep increase of $c_2$ around $T_c$ we need to introduce hadronic contributions up to right above $T_c$. At even larger temperature the hadrons seem to be replaced by a almost non interacting gas of quarks.
This offers an intriguing implication concerning the CeP. In mean field studies with the PNJL \cite{Fukushima:2008is} as well as in the HQ model \cite{Steinheimer:2010ib}, a general feature of including repulsive interactions is that the CeP is moved to larger chemical potentials or even disappears completely. From this point of view, a vanishing quark vector interaction would favor the existence of a CeP.\\
On the other hand, the phase structure of QCD up to $T_c$ and likely even slightly above it could be determined by hadronic interactions. In model calculations without any repulsive quark interactions the quarks often appear at rather small chemical potentials (at $T=0$) making a reasonable description of the nuclear ground state difficult. A repulsive hadronic vector interaction, which is required on order to reproduce the properties of a saturated nuclear ground state, may move the CeP to larger chemical potentials or even remove it completely from the phase diagram in accordance to mean-field results \cite{Fukushima:2008is,Ferroni:2010xf}.
\section{Acknowledgments}
This work was supported by BMBF, GSI and the Hessian LOEWE initiative through the Helmholtz International center for FAIR (HIC for FAIR). The computational resources were provided by the Frankfurt Center for Scientific Computing (CSC).\\

\end{document}